\documentclass[a4paper]{jpconf}
\pdfoutput=1
\usepackage{graphicx}
\usepackage{cite}
\usepackage{amsmath}
\usepackage{amssymb}
\usepackage{multirow}
\usepackage[T1]{fontenc}

\usepackage[pdftex,usenames,dvipsnames]{xcolor}
\usepackage[bookmarks]{hyperref}
        \hypersetup{colorlinks,
                    linktocpage,
                    citecolor=MidnightBlue,
                    filecolor=Blue,
                    linkcolor=MidnightBlue,
                    urlcolor=MidnightBlue}
\newcommand {\pp}    {pp}

\newcommand {\PbPb}  {Pb--Pb}

\newcommand {\sqrtS}   {\mbox{$\sqrt{s}$}}
\newcommand {\sqrtSnn} {\mbox{$\sqrt{s_{\text{\textsc{nn}}}}$}}
\newcommand {\vTwo}    {\mbox{$v_{\mathrm{2}}$}}
\newcommand {\Raa}     {\mbox{$R_{\mathrm{AA}}$}}
\newcommand {\Pt}      {\ensuremath{p_{\text{\textsc{t}}}}}
\newcommand {\Taa}     {\mbox{$\langle T_{\mathrm{AA}} \rangle$}}
\newcommand {\Npart}   {\mbox{$\langle N_{\mathrm{part}} \rangle$}}
\newcommand {\YieldAA} {\mbox{$\mathrm{d}N_{\mathrm{AA}}/\mathrm{d}\Pt$}}
\newcommand {\xSecpp}  {\mbox{$\mathrm{d}\sigma_{\mathrm{pp}}/\mathrm{d}\Pt$}}
\newcommand {\GeVc}    {\mbox{GeV/$c$}}

\newcommand {\Dplus} {\mbox{$\mathrm{D^{+}}$}}
\newcommand {\Ds}    {\mbox{$\mathrm{D^{+}_{s}}$}}
\newcommand {\Dzero} {\mbox{$\mathrm{D^{0}}$}}
\newcommand {\Dstar} {\mbox{$\mathrm{D^{*+}}$}}
\newcommand {\Jpsi}  {\mbox{$\mathrm{J/\Psi}$}}

\begin{document}
      \title{Open-charm measurements in \pp{} and \PbPb{} collisions at central rapidity with ALICE}

      \author{Julien Hamon, for the ALICE collaboration}
      \address{Universit\'e de Strasbourg, CNRS-IN2P3, IPHC UMR 7178, F-67000 Strasbourg, France}
      \ead{julien.hamon@iphc.cnrs.fr}

      \begin{abstract}
An overview of recent \Dplus{}, \Dzero{}, \Dstar{} and \Ds{} measurements performed by ALICE at central rapidity in proton-proton collisions at $\sqrtS=7$ and 8~TeV, as well as lead-lead collisions at $\sqrtSnn=2.76$ TeV, is presented. An emphasis is put on the discussion of theoretical predictions with respect to the LHC Run I data.

\smallskip \noindent
Presented at the conference "Hot Quarks 2016", South Padre Island, Texas, USA. September 12--17, 2016.
{\scriptsize Link to the presentation: \url{https://indico.cern.ch/event/507867/contributions/2218105/}}
      \end{abstract}

      \section{Introduction}

Heavy quarks (charm and beauty) are unique probes to investigate the properties of the hot and dense matter created in ultra-relativistic heavy-ion collisions, known as the Quark-Gluon Plasma. Due to their large masses, heavy quarks are mainly produced at the early stage of the collision in hard parton scatterings, before the plasma formation, and thus participate in all the subsequent stages of the collision.
In their propagation through the medium, they interact with other partons and lose energy via elastic (collisional) and inelastic (gluon radiation) processes, commonly treated in the perturbative approach of Quantum Chromo-Dynamics (QCD). These processes are expected to depend on the parton in-medium path length, colour-charge and mass \cite{bdmps1997radiativeEnergyLosses,dokshitzer2001heavyQuarksCalorimetry}. This results in a characteristic ordering of the energy loss of partons:
$\Delta E(\mathrm{g}) > \Delta E(\mathrm{u,d,s}) > \Delta E(\mathrm{c}) > \Delta E(\mathrm{b})$.

Experimentally, heavy-flavour hadrons are the front door to access medium properties probed by heavy quarks. During the LHC Run I, ALICE \cite{alice2008pstationExpmtInJINST} has tackled the measurements of charm and beauty hadrons in several rapidity ranges. At central rapidity ($|y_{\text{lab}}|<0.5$), open-charm mesons have been studied in various colliding systems and energies. The tracking and particle identification abilities of the ALICE central barrel detectors have allowed for measuring \Dplus{}, \Dzero{}, \Dstar{} and \Ds{} over a wide transverse momentum (\Pt{}) range and for different collision centralities.

      \section{\pp{} collisions at $\sqrtS = 7$ and $8$ TeV}

Proton-proton (\pp{}) collisions provide a suitable environment for testing several predictions of QCD. The large mass of charm quark allows for perturbative calculations of the differential charm production down to low \Pt{}, as well as its total production cross section. At LHC energies, the study of small systems also permits to examine the gluon and heavy-quark distributions inside the nucleon. Besides, \pp{} collisions represent a crucial baseline for understanding the phenomena observed in heavy-ion collisions, by acting as a reference where final state effects are negligible.

\begin{figure}[h]
         \centering
         \begin{minipage}{14pc}
            \hspace{-1.4cm}
            \includegraphics[scale=0.37]{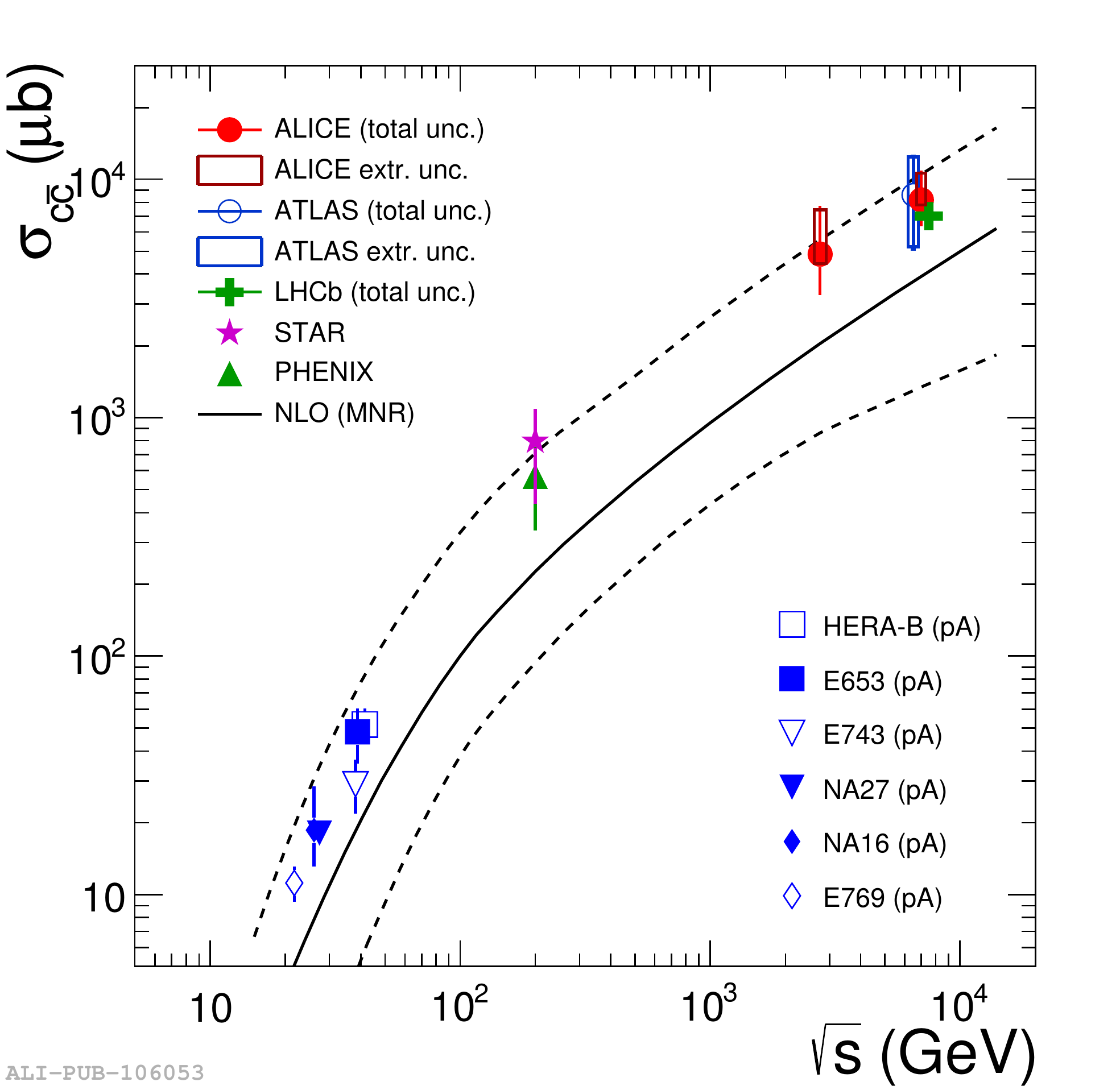}
         \end{minipage}\hspace{3pc}
         \begin{minipage}{14pc}
            \vspace{-0.26cm}
            \includegraphics[scale=0.37]{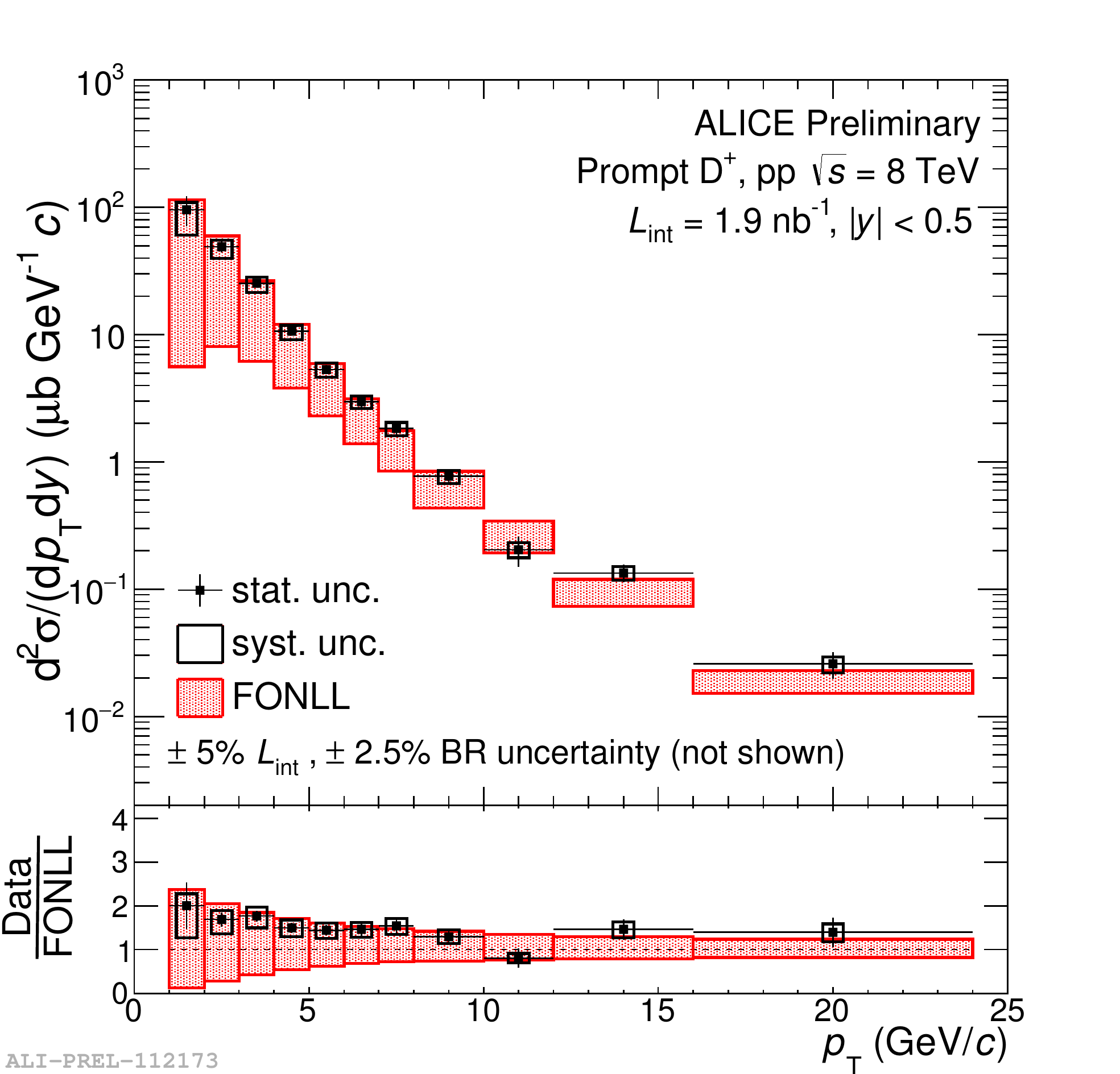}
         \end{minipage}
         \caption{Left: Total charm production cross section in nucleon-nucleon collisions as a function of \sqrtS{} \cite{alice2016DmesonProductionInpPb5pt02TeVAndppAt7TeV}. Right: Preliminary \Pt{}-differential production cross section of prompt \Dplus{} in \pp{} collisions at $\sqrtS=8$~TeV, measured by ALICE.}
         \label{fig:pp8TeV}
      \end{figure}

In this context, the \Pt{}-differential production cross section of prompt \Dzero{} has been recently measured down to $\Pt{}=0$ \GeVc{} in pp collisions at $\sqrtS = 7$ TeV by ALICE \cite{alice2016DmesonProductionInpPb5pt02TeVAndppAt7TeV}, with the help of an analysis method which does not exploit the D-meson decay (displaced) vertex. Several pQCD models based on a factorisation theorem (FONLL \cite{fonll2012CharmAndBottomProductionAtLHC}, GM-VFNS \cite{gmvfns2012CharmProductionAtLHC}, $k_{\text{\textsc{t}}}$-factorisation \cite{kTfactorisation2013CharmProductionAtLHC}) are in agreement with this measurement within the current theoretical precisions.
A similar agreement is achieved for all other D-meson species.
Furthermore, the total charm cross section at $\sqrtS=7$ TeV has been updated and is well reproduced by pQCD calculations at next-to-leading order \cite{mnr1992totalCharmCrossSectionAtNLO}, within quite large theoretical uncertainties though, as presented in figure \ref{fig:pp8TeV} (left).

Currently, the production cross sections of prompt \Dplus{} and \Dstar{} as a function of \Pt{} are being studied by ALICE in pp collisions at $\sqrtS = 8$ TeV. The \Dplus{} preliminary results are shown in figure \ref{fig:pp8TeV} (right). As already noticed at lower energies, the FONLL pQCD framework describes the data reasonably well within uncertainties, although the data still sit at the upper edge of the calculations. As expected, the production cross section obtained at 8 TeV is slightly higher than the one at 7 TeV.

      \section{\PbPb{} collisions at $\sqrtSnn=2.76$ TeV}

The parton energy loss, expected for heavy quarks in nucleus-nucleus collisions, leads to a reduction of the average parton momentum, which can be experimentally studied -- to some extent -- through the measurement of the nuclear modification factor of open-charm hadrons:
   \begin{equation}
      \Raa (\Pt) = \frac{1}{\Taa} \cdot \frac{ \YieldAA }{ \xSecpp }
   \end{equation}
where \Taa{} is the average nuclear overlap function, proportional to the average number of nucleon-nucleon binary collisions, \YieldAA{} and \xSecpp{} the transverse momentum differential yield and cross section in nucleus-nucleus and proton-proton collisions, respectively.

\begin{figure}[h]
         \centering
         \begin{minipage}{14pc}
            \hspace{-2.0cm}
            \includegraphics[scale=0.35]{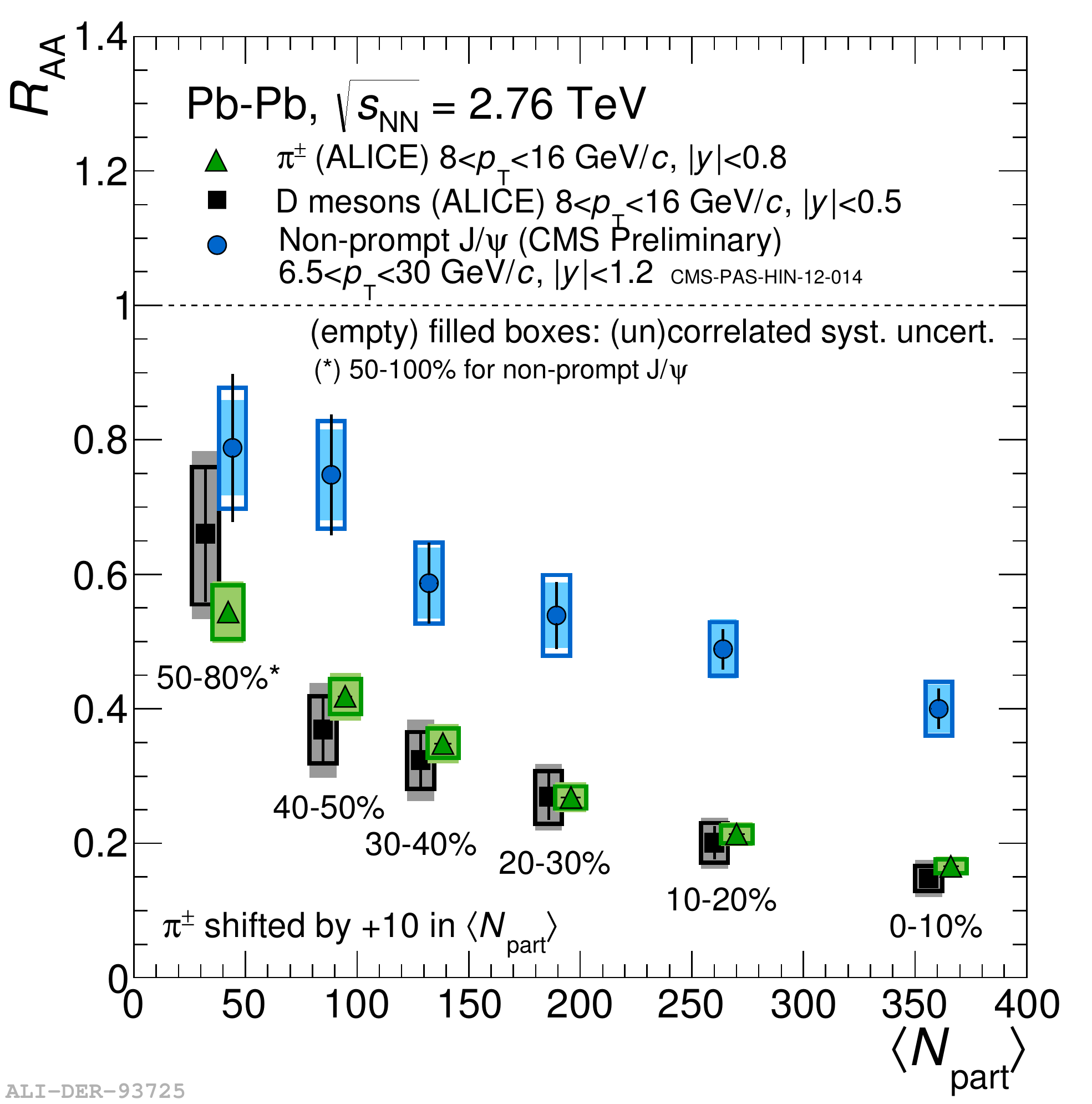}
         \end{minipage}\hspace{1pc}%
         \begin{minipage}{14pc}
            \includegraphics[scale=0.37]{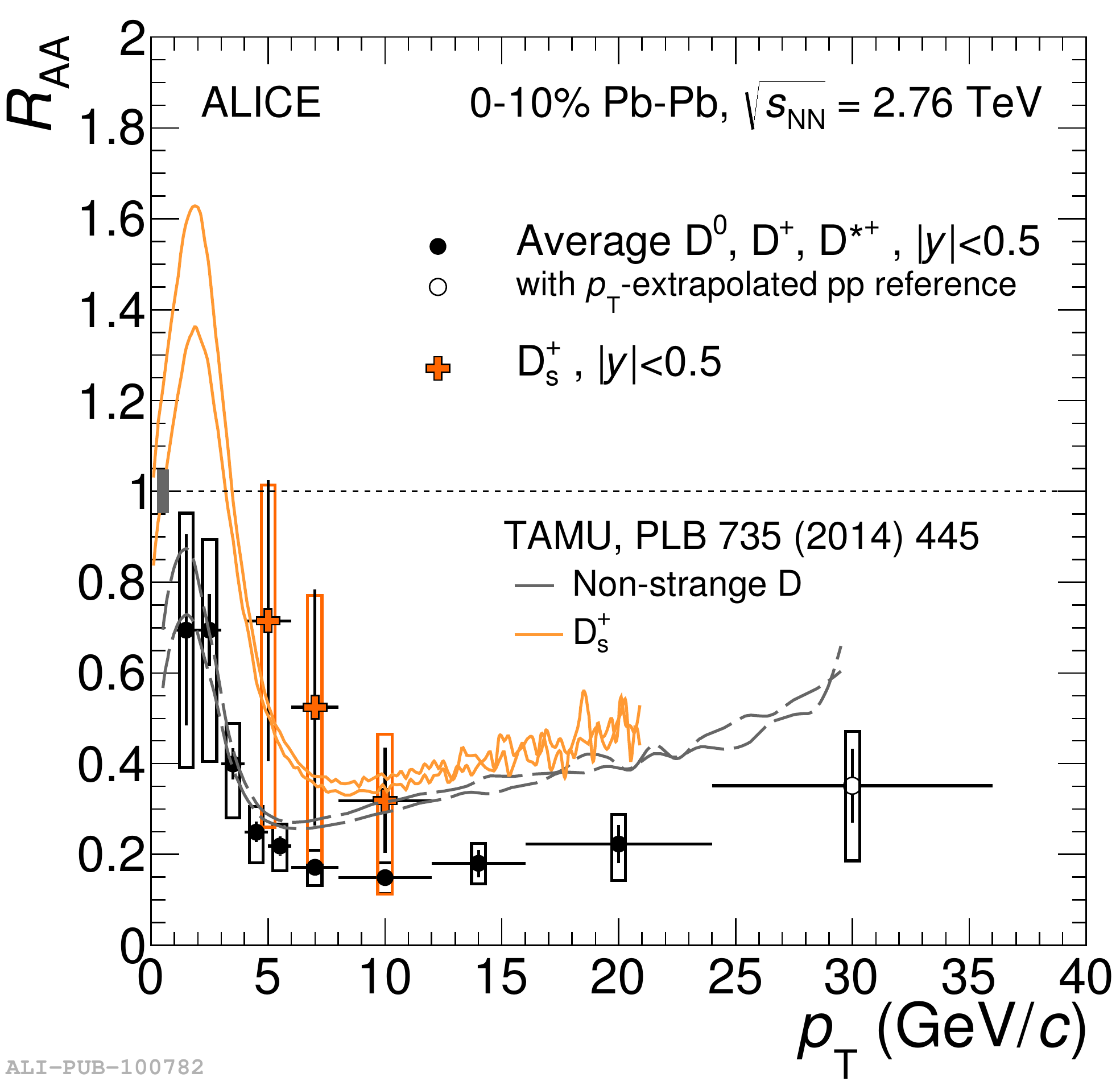}
         \end{minipage}
         \caption{Nuclear modification factor of non-strange D mesons as a function of \Npart{} \cite{alice2015centralityDependenceOfHighPtDmesonsInPbPbAt2pt76TeV}, compared to that of charged pions \cite{alice2012suppressionHighPtDmesonsInCentralPbPbAt2pt76TeV} and non-prompt \Jpsi{} \cite{cmsUptade2012CentralityDependenceOfJpsiInPbPb2pt76TeV} (left); and as function of \Pt{} compared to that of \Ds{} for central \PbPb{} collisions at $\sqrtSnn=2.76$ TeV \cite{alice2016PtDependenceOfDmesonProdInPbPbAt2pt76TeV,alice2016DsProductionAndRaaInPbPbAt2pt76TeV} (right).}
         \label{fig:PbPbResults}
      \end{figure}

ALICE has extensively studied the nuclear modification factor of D mesons in lead-lead (\PbPb{}) collisions at $\sqrtSnn = 2.76$ TeV \cite{alice2012suppressionHighPtDmesonsInCentralPbPbAt2pt76TeV,alice2016PtDependenceOfDmesonProdInPbPbAt2pt76TeV,alice2016DsProductionAndRaaInPbPbAt2pt76TeV}. The observed stronger high-\Pt{} suppression going from semi-central collisions ($\Raa(\Pt\sim6~\GeVc)\geq0.4$) to central ones ($\Raa(\Pt\sim6~\GeVc)\leq0.2$) is interpreted as an increase of the medium density, size and lifetime leading to larger energy loss of charm quarks.
Additionally, the nuclear modification factor of D mesons has been measured as a function of \Npart{}, the average number of nucleons participating in the \PbPb{} collisions \cite{alice2015centralityDependenceOfHighPtDmesonsInPbPbAt2pt76TeV}. The results are shown in figure \ref{fig:PbPbResults} (left), where they are compared to those obtained by ALICE for charged pions \cite{alice2014productionPionsKaonsProtonsAtLargePtInppAndPbPbAt2pt76TeV} and by CMS for \Jpsi{} coming from beauty decays \cite{cmsUptade2012CentralityDependenceOfJpsiInPbPb2pt76TeV}, in a similar kinematic range (high \Pt{} and central rapidity).
As can be seen, $\Raa(\Jpsi\leftarrow\mathrm{B})$ is found to be significantly larger than $\Raa(\Dplus,\Dzero,\Dstar)$ in central collisions, which is consistent with the parton energy loss hierarchy. Such an observation is in qualitative agreement with several theoretical predictions involving a mass dependence of parton energy loss processes (Djordjevic \emph{et al.} \cite{djordjevic2014RHICandLHCjetSuppressionInNonCentralCollisions}, MC$@$sHQ$+$EPOS2 \cite{mcshqepos2014heavyQuarksInPbPbAtLHC} and TAMU elastic \cite{tamu2013heavyFlavourTransportAtRHICandLHC}). On the other hand, the data show a quite similar \Raa{} for high-\Pt{} pions and D mesons. This might be explained as due to the interplay between the colour-charge dependence of parton energy loss, and the different \Pt{} and fragmentation distributions for gluons and charm quarks. At low \Pt{}, any direct interpretation is complicated since, in contrast to heavy-flavour hadrons, light-flavour hadron yields do not scale with the number of nucleon-nucleon binary collisions but rather with the number of participant nucleons. Moreover, radial flow and cold nuclear matter effects affect differently pions and D mesons.

The first measurement of the \Ds{} nuclear modification factor has been performed by ALICE in \PbPb{} collisions at $\sqrtSnn = 2.76$ TeV \cite{alice2016DsProductionAndRaaInPbPbAt2pt76TeV} and is presented in figure \ref{fig:PbPbResults} (right). Although $\Raa(\Ds)$ is compatible with the average \Raa{} of non-strange D mesons within uncertainties, the current precision does not allow for drawing a firm conclusion on a possible signature of recombination mechanism working for charm quarks. An increase of the relative fraction of strange versus non-strange D mesons is indeed expected by the TAMU model, at low and intermediate \Pt{} \cite{tamu2014heavyFlavorAtLHC}.

The diversity of open-charm measurements carried out by ALICE offers a valuable basis to constrain models. The results encompass both nuclear modification factor and elliptic flow \cite{alice2014DmesonFlowAndInPlaneRAAInPbPb2pt76TeV} of different D-meson species as a function of \Pt{} and the collision centrality. An attempt to describe simultaneously all measured observables, for charged pions and D mesons, within various theoretical models has been made in a recent ALICE paper \cite{alice2016PtDependenceOfDmesonProdInPbPbAt2pt76TeV}.
While all models rely on a colour-charge and mass dependence of energy loss, the interaction mechanisms are not yet clearly accessed. As a general rule, gluon radiation seems to be needed to describe high-\Pt{} \Raa{}(D), whereas collisional interactions appear to be important for the understanding of low-\Pt{} \Raa{} and \vTwo{} measurements. Although the role of the  recombination -- which should help describing low-\Pt{} \vTwo{} -- is open to debate, it is generally accepted that an expanding medium gives rise to the observed non-zero \vTwo{}.
As shown in \cite{alice2016PtDependenceOfDmesonProdInPbPbAt2pt76TeV}, models fairly describing heavy-flavour hadron data do not necessarily manage to simultaneously reproduce pion data. In particular, only few models are able to describe both hadron species but only at high \Pt{}: Djordjevic \emph{et al.} \cite{djordjevic2014RHICandLHCjetSuppressionInNonCentralCollisions}, CUJET$3.0$ \cite{gyulassy2015cujet3} and Vitev \emph{et al.} \cite{vitev2009inMediumDissociation}. It is worth noting that the latest includes in-medium meson formation and dissociation as an energy loss mechanism, on top of collisional interactions, instead of the commonly used radiative process.

      \section{Summary}

Charm quark is an efficient tool to probe QCD in its perturbative regime and to give insights into the various stages of a heavy-ion collision, from its early stage to the interaction processes occurring in the hot and dense medium until hadronization. By exploiting LHC Run I, ALICE started providing constraints on theoretical models aiming at the description of charm physics. Energy loss processes can be studied with the diversity of measured observables. However, a simultaneous description of all open-charm data stays challenging. More sophisticated models together with more precise measurements, coming from the ongoing LHC Run II data taking, will be further needed to draw a clearer picture of the interaction mechanisms taking place throughout heavy-ion collisions.

      \section*{References}

\bibliographystyle{myJHEP}
\bibliography{BibTex_MyBibliography}

\end{document}